\documentclass[journal]{IEEEtran}

\usepackage{amsfonts,graphicx,amsmath,latexsym,amssymb,enumerate,cite,color,bm}
\usepackage{subfigure}
\usepackage{caption}
\usepackage{comment}
\newcommand{\R}{\mathbb{R}}

\newcommand{\ddd}{\mathsf{d}}

\newcommand{\ccc}{\mathsf{c}}

\definecolor{GREEN}{rgb}{0.0, 0.5, 0.0}

\newtheorem{prop}{\bf Proposition}
\newtheorem{defn}{\bf Definition}
\newtheorem{lem}{\bf Lemma}
\newtheorem{thm}{\bf Theorem}
\newtheorem{asm}{\bf Assumption}
\newtheorem{rem}{\bf Remark}

\begin{document}

\title{A Zero-stealthy Attack for Sampled-data Control Systems via Input Redundancy}

\author{Jihan Kim,
        Gyunghoon Park,
        Hyungbo Shim, {\it Senior Member, IEEE}, and 
        Yongsoon Eun, {\it Member, IEEE}
\thanks{The material in this paper was partially presented at the 55th IEEE Conference on Decision and Control (CDC), December 12--14, 2016, Las Vegas, NV, USA \cite{KPSE16}.}
\thanks{This work was supported by Institute for Information \& communications Technology Promotion(IITP) grant funded by the Korea government(MSIT) (2014-0-00065, Resilient Cyber-Physical Systems Research).}
\thanks{J.~Kim, G.~Park, and H.~Shim are with ASRI, Department of Electrical and Computer Engineering, Seoul National University, Korea. Y.~Eun is with Department of Information \& Communication Engineering, Daegu Gyeongbuk Institute of Science \& Technology, Korea.}
}

\markboth{}
{Shell \MakeLowercase{\textit{et al.}}: Bare Demo of IEEEtran.cls for IEEE Journals}

\maketitle

\begin{abstract}
	In this paper, we introduce a new vulnerability of cyber-physical systems to malicious attack.
	It arises when the physical plant, that is modeled as a continuous-time LTI system, is controlled by a digital controller.
	In the sampled-data framework, most anomaly detectors monitor the plant's output only at discrete time instants, and thus, nothing abnormal can be detected as long as the sampled output behaves normal.
	This implies that if an actuator attack drives the plant's state to pass through the kernel of the output matrix at each sensing time, then the attack compromises the system while remaining stealthy.
	We show that this type of attack always exists when the sampled-data system has an input redundancy, i.e., the number of inputs being larger than that of the outputs or the sampling rate of the actuators being higher than that of the sensors.
	Simulation results for the X-38 vehicle and for the other numerical examples illustrate this new attack strategy possibly brings disastrous consequences.
\end{abstract}


\IEEEpeerreviewmaketitle

\section{Introduction}

\IEEEPARstart{R}{ecent} development of communication capabilities and computational resources has led to the integration of cyber-technologies and physical processes, which improves efficiency and flexibility of the system. 
These Cyber-Physical Systems (CPS) include not only simple or small devices, but also a variety of critical infrastructures that are closely related to public health and numerous financial costs. 
Examples include nuclear facilities, power grid (smart grid), supervisory control and data acquisition (SCADA) system, and networked transportation. 
For this reason, the security problem of CPS has received a lot of attention in recent years.

In particular, cyber-attacks on CPS may bring disastrous consequences, and their impacts are well illustrated by subsequent incidents, such as the Stuxnet attack on Iran's nuclear plant \cite{JR11}, massive power blackouts in South America \cite{J10}, Maroochy water breach in Australia \cite{JM07}, and cyber-attack on the Ukrainian power grid \cite{LAC16}. These instances highlight the need for measuring the vulnerabilities of CPS against malicious attacks and unexpected errors. There have been several researches that examine the vulnerabilities of CPS from the control-theoretic point of view. For instance, the weakness of electric power grids, possibly caused by false data injection attacks, was studied in \cite{YP11}. An undetectable sensor attack to the unstable system was presented in \cite{YB10}. More recently, the authors of \cite{AIHK15} explored the question which resources should be utilized for the attack design, also focusing on various attack scenarios including denial of service (DoS) attack \cite{SAS09}, replay attack \cite{YB09}, zero-dynamics attack \cite{AIHK12}, local zero-dynamics attack, and bias injection attack. 

It is worth mentioning that most of the researches on security problems of CPS have been studied either in continuous-time or in discrete-time domain. 
From a practical standpoint, however, usual cyber-physical systems are composed of continuous-time physical plants and discrete-time digital controllers. 
It means that, for thorough understanding of cyber-security, interaction between the continuous-time and discrete-time components should come into the picture.
In this regard, we are concerned with the security problem for sampled-data control system that consists of a multi-input multi-output (MIMO) continuous-time plant, samplers, and zero-order hold (ZOH) devices.
Specifically, we allow that the sampling rate of the actuators be different from that of the sensors. 
These multi-rate sampling schemes have been widely studied in the literature for specific purposes. 
For example, a faster actuation than sensing has been adopted to improve control performance such as inter-sample behavior, disturbance rejection, and so on \cite{LW00,HY02,HFS03,KB08,MS07}.
On the other hand, faster sensing has advantages on state feedback control design, acceleration control, and security problem  \cite{MNP15,HA1988,MTO2007}.

In this paper, we show that the sampled-data systems are possibly vulnerable to a malicious adversary who utilizes an {\em input redundancy} of the systems.
This redundancy becomes available to the attacker when (a) the sampling rate of the actuator is faster than that of the sensor, or (b) the number of inputs is larger than that of the outputs. 
Using the input redundancy, we present a new type of stealthy attack in the sampled-data framework. 
The underlying idea for the attack design is to express the sampled-data system as an extended {\em lifted system} with a stacked state variable, and to enforce the state to remain a (nontrivial) kernel of its output matrix (at sampling times).
In doing so, the attack cannot be detected by any (discrete-time) anomaly detector that is built upon the sampled measurements of the output;
at the same time, the inter-sample behavior of the physical plant is compromised.
We will show that all of these can be done with the input redundancy. 
It should be pointed out that, unlike the well-known zero-dynamics attack \cite{AIHK12,AIHK15,Park16}, the proposed attack policy is applicable even when there is no unstable zero (either for continuous-time model or for its sampled-data counterpart).

The remainder of this paper is organized as follows. 
Section \ref{sec: Problem Formulation} presents the problem formulation. 
Section \ref{sec: main result} provides an attack design and studies when and how the adversary successfully spoils the sampled-data control systems.
A few numerical examples and case studies can be found in Section \ref{sec:Sim}.
Concluding remarks and further discussions are given in Section \ref{sec: conclusion}.

{\em Notation}: For two vectors $a$ and $b$, ${\rm col}(a,b)$ stands for $[a^T~b^T]^T$. The sets of natural, rational, and real numbers are denoted by $\mathbb{N}, \mathbb{Q},$ and $\mathbb{R}$, respectively. The notation $\|x\|$ denotes the Euclidean norm for vector $x$. For a real number $r\in\mathbb{R}$, $\lfloor r \rfloor$ denotes the largest integer which is smaller than or equal to $r$.
For a matrix $A$, $\ker A$ implies the null space of $A$ and ${\rm im}\; A$ is the range space of $A$.

\section{Problem Formulation}\label{sec: Problem Formulation}

We consider a compromised continuous-time physical system modeled as
\begin{equation}
\begin{aligned}
\dot{x}(t) &= Ax(t)+B(u(t)+a(t)), \\ 
y(t) &= Cx(t)
\end{aligned}\label{ctsys}
\end{equation}
where $x \in \mathbb{R}^n$ is the system state, $u, a \in \mathbb{R}^p$, and $y\in \mathbb{R}^q$ are the input, a malicious attack, and the output of the system, respectively, and $A\in \mathbb{R}^{n\times n}, B\in \mathbb{R}^{n\times p}, $ and $C\in \mathbb{R}^{q\times n}$.

\begin{figure}[t!]
	\vspace{0.2cm}
	\begin{center}
		\includegraphics[width=0.45\textwidth]{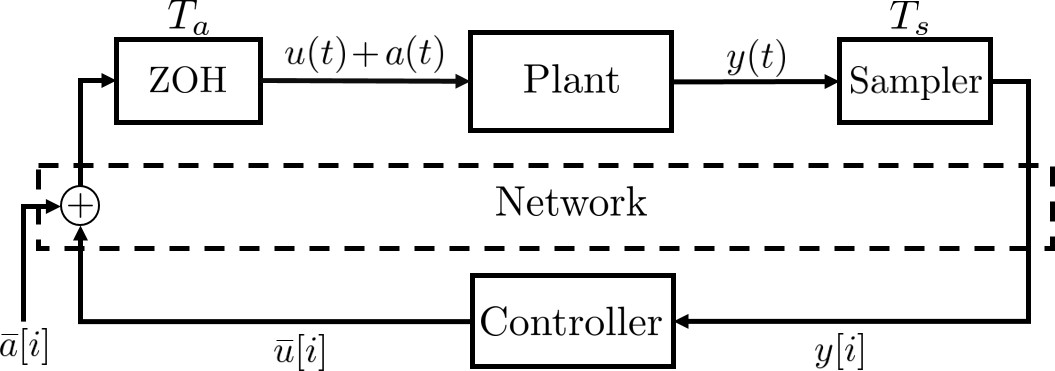}
	\end{center}
	\caption{Sampled-data system connected through network}\label{fig:SDsys}
\end{figure}

Throughout the paper we suppose that the plant \eqref{ctsys} is connected with a discrete-time controller through a communication network as seen in Fig.~\ref{fig:SDsys}.
Specifically, it is assumed that the discrete-time control is performed with the ``sampler'' for the output $y(t)$ with the sampling period $T_s$, and the ``zero-order holder (ZOH)'' for both the input $u(t)$ and the attack signal $a(t)$ with the sampling period $T_a$.
Hence, $u(t)$ and $a(t)$ are piecewise constant functions such that $u(t) = u(iT_a)$ and $a(t)=a(iT_a)$ for $iT_a \leq t < (i+1)T_a$.
It is supposed that $u(iT_a) = \bar u[i]$ and $a(iT_a)=\bar a [i]$ where $\bar u[i]$ is the output of a discrete-time controller and $\bar a [i]$ is a discrete-time attack signal injected through the vulnerable input communication network.

\begin{figure}[t!]
  	\begin{center}
  		\includegraphics[width=0.45\textwidth]{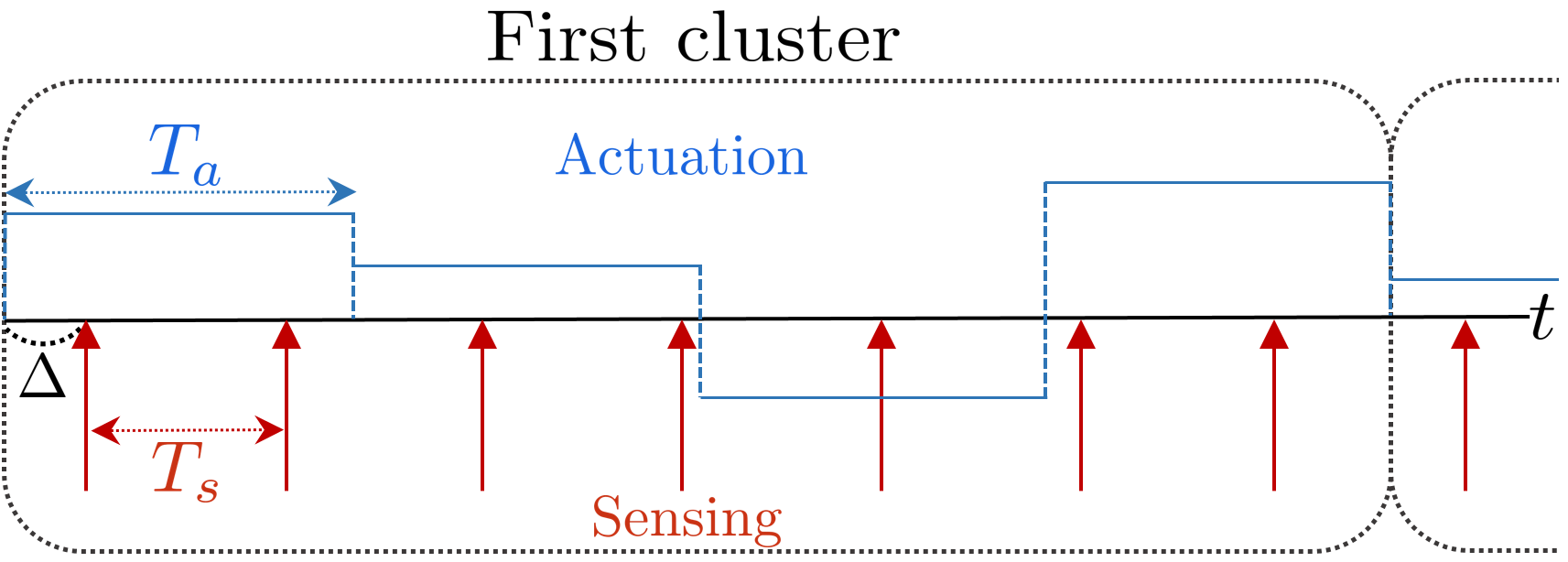}
  	\end{center}
  	\caption{Example of a cluster when $R=T_s/T_a=4/7=\beta/\alpha $. There are $\beta$ actuation times and $\alpha$ sensing times in one cluster.}\label{fig:cluster_S1}
  \end{figure}

In this paper, we are interested in general multi-rate sampled-data systems where $T_s$ and $T_a$ are not necessarily the same.
In particular, the ratio between $T_s$ and $T_a$ is assumed to satisfy 
\begin{equation}
R := \frac{T_s}{T_a} \in \mathbb{Q}. \label{ratio_N}
\end{equation}
In what follows, we often use the coprime fraction $R =\beta/\alpha$ with $\alpha, \beta \in \mathbb{N}$ (rather than \eqref{ratio_N}). 
It should be noticed that, while the actuation times are $t=iT_a$ with $i=0, 1, \cdots$, there is no reason that the sensing time (when the output $y(t)$ is sampled) is synchronized with the actuation time in practice.\footnote{Refer to \cite{KPSE16} for synchronous case. While \cite{KPSE16} has more limitations such as $R$ being integer, the derivation of \cite{KPSE16} is simpler than this paper.}
So, let us suppose that the sensing times are $t=jT_s + \Delta$ with $j=0, 1, \cdots$, where $0 \le \Delta < T_s$ is called an {\em offset} in this paper.
Note that, while actuation times and sensing times are asynchronous, distribution of their times exhibits a pattern that repeats in every $\alpha T_s = \beta T_a$ seconds (see Fig.~\ref{fig:cluster_S1}) since $R=T_s/T_a=\beta/\alpha$.

For convenience, we define a normalized offset $\delta :=\Delta/T_s$ (so that $0 \leq \delta < 1$), and a new index (which is a real number) as
$$j_{\delta} := \delta + \lfloor j-\delta \rfloor, \qquad j = 1, 2, \cdots.$$ 
Then $j_\delta = (j-1)+\delta$ if $\delta > 0$ and $j_\delta = j$ if $\delta =0$.
Using the index $j_\delta$, the sampled-data system {\em in terms of the sensing times} can be written in the discrete-time domain as
\begin{align}
 x((j_{\delta}+1) T_s) &= e^{AT_s}x(j_{\delta} T_s)  \notag \\ 
 &\hspace{-4mm} + \int_{j_{\delta} T_s}^{(j_{\delta}+1)T_s} e^{A((j_{\delta} +1)T_s-\tau)} B (u(\tau) + a(\tau))  d\tau, \notag \\
 y(j_\delta T_s) &= Cx(j_\delta T_s) =: \bar y[j] \label{dtsys}
\end{align} 
for $j=1,2,\cdots$, while $x(1_\delta T_s)$ is given by $x(1_\delta T_s) = e^{A 1_\delta T_s} x(0) + \int_{0}^{1_\delta T_s} e^{A(1_\delta T_s-\tau)} B (u(\tau)+a(\tau)) d\tau$.

Without loss of generality, it is assumed that $t=0$ be the time when the attack is initiated.
Now, for comparison, let $x_{\mathsf{o}}$ be the solution of \eqref{dtsys} without any attack (i.e., $a(t) \equiv 0$) and let $y_{\mathsf{o}} = Cx_{\mathsf{o}}$.
It is noted that $x_{\mathsf{o}}(0)=x(0)$ since the attack starts at $t=0$.
Then, with the error variables
$$\tilde{x}(t):=x(t)-x_{\mathsf{o}}(t),\quad \tilde{y}(t):=C\tilde{x}(t),$$
we have the error dynamics (obtained from \eqref{dtsys}) as
\begin{align}
 \tilde{x}((j_{\delta}+1) T_s) & = e^{AT_s}\tilde{x}(j_{\delta} T_s) \notag\\
 &\quad +\int_{j_{\delta} T_s}^{(j_{\delta}+1)T_s}e^{A((j_{\delta}+1)T_s-\tau)} B a(\tau) d\tau,\notag\\
 \tilde{y}(j_{\delta} T_s)&=C\tilde{x}(j_{\delta} T_s) \label{xasys}
\end{align}
with $\tilde x(1_\delta T_s) = \int_0^{1_\delta T_s} e^{A(1_\delta T_s - \tau)} B a(\tau) d\tau$.

The problem to be studied is to generate an attack signal $\bar a[i]$ having the following two important features simultaneously. 

\begin{defn}
	An attack sequence $\{\bar{a}[i]\}_{i=0}^{\infty}$ is said to have {\it zero-stealthy property} if $\tilde{y}(j_\delta T_s)\equiv0$ for all $j\geq 1$. \hfill$\square$
\end{defn}

This property directly implies that $y(j_\delta T_s)\equiv y_{\mathsf{o}}(j_\delta T_s)$ for all $j\geq 1$, and thus, the plant \eqref{dtsys} under the attack seemingly operates normally as if it is attack-free.
Thus, no anomaly detector that uses $\bar u[i]$ and $\bar y[j]$ can detect the attack.

\begin{defn}
	For a given sequence of positive thresholds $\{ H_k \}_{k=1}^{\infty}$, an attack sequence $\{\bar{a}[i]\}_{i=0}^{\infty}$ is said to have {\it disruptive property} with $\{ H_k \}_{k=1}^{\infty}$, if $\left\|\tilde{x}(t_k)\right\| \ge H_k$ for all $k \ge 1$ with a time sequence $\{t_k\}_{k=1}^{\infty}$ satisfying $(k-1)(\beta T_a) < t_k \leq k (\beta T_a)$. \hfill $\square$
\end{defn}

The {\it disruptive property} indicates that the size of the error state $\tilde x(t)$ becomes larger than $H_k$ at least once within the $k$-th time interval of the length $\beta T_a$ (that is, the time interval for $\beta$ times of actuations, or $\alpha$ times of measurements).
Strength of the attack can be considered as the values of the sequence $\{ H_k \}_{k=1}^{\infty}$, whose selection is fully upon the adversary.

A conventional solution to this problem is, as widely studied in the literature, the so-called zero-dynamics attack \cite{AIHK12}.
However, this attack is effective only when the system is of non-minimum phase, and the strength of the attack is determined solely by the plant's zero-dynamics, and so the attacker is not able to assign the speed of divergence.
Moreover, it is not a completely stealthy attack in the sense that its initiation causes a transient that can be observed from the output. 
(Therefore, in practice, the initial condition of the zero-dynamics attacker is set to be small enough so that the transient can hide below the alarm level in the anomaly detector.)
On the other hand, the proposed attack is `zero-stealthy' implying that the attacked output is {\it exactly} the same as the normal one at every sampling times.

In this paper, we propose a zero-stealthy disruptive attack for the sampled-data system that is possibly more lethal than the conventional zero-dynamics attack. 
Our proposal is based on the assumption that the sampled-data system \eqref{dtsys} has a kind of {\it input redundancy}. 
This is the case when the zero-order holder works faster than the sampler (that is, $R$ is larger than 1), or the number $p$ of the input channel is larger than that of the output channel, $q$. Then, as we shall see below, the adversary can generate a new type of stealthy attack that has disruptive behavior with arbitrarily large thresholds.

\section{Design of Zero-stealthy Attack with Disruptive Property}\label{sec: main result}

The first task for the attack design is to rewrite the sampled-data system \eqref{xasys} in the actuation time frame with $T_a$ as 
\begin{align}\label{xa dynamics}
\begin{split}
\tilde{x}(iT_a) &= A_\ddd^i \tilde{x}(0)+\sum\limits_{m=0}^{i-1}A_\ddd^{i-1-m}B_\ddd\bar a[m] \\
&= \sum\limits_{m=0}^{i-1}A_\ddd^{i-1-m}B_\ddd\bar a[m]
\end{split}
\end{align}
where the last equality follows from $\tilde x (0)=0$, and 
\begin{equation}
A_\ddd := e^{A T_a} \in \mathbb{R}^{n\times n}, ~ B_\ddd := \bigg(\int^{T_a}_0 e^{A\tau}d\tau\bigg)B \in \mathbb{R}^{n\times p}.\notag
\end{equation}
For progression, we need generalized notations about $A_\ddd$ and $B_\ddd$, which are related to both $T_s$ and $T_a$ as follows:
$$A_\ddd^{\langle l,m \rangle}:=e^{A(lT_s-mT_a)}, \;\;  B_\ddd^{\langle l,m \rangle} := \bigg(\int_{0}^{lT_s-mT_a}e^{A\tau}d\tau\bigg) B.$$ 
From the above definition, $A_\ddd$ and $B_\ddd$ also can be denoted as $A_\ddd^{\langle 0,-1 \rangle}$ and $B_\ddd^{\langle 0,-1 \rangle}$, respectively.

\subsection{Clustering the Time Frame}
  
To construct the attack sequence $\bar a [i]$ efficiently, let us introduce the concept of `cluster.' 
The $k$-th cluster is defined as the time period $(k-1)\beta T_a < t \le k\beta T_a$ (and sometimes we indicate the left-closed interval $(k-1)\beta T_a \le t < k\beta T_a$ by calling it the $k$-th {\em input-}cluster).
It can be seen that each cluster contains exactly $\alpha$ sensing times and $\beta$ actuation times.
See Fig.~\ref{fig:cluster_S1} for the case of $R=4/7$ with $\Delta >0$.
By exploiting these clusters, we will consider the error dynamics \eqref{xasys} in terms of the clusters.
For this, let us define stacked attack vector in the $k$-th input-cluster, and the stacked states and the stacked measurements in the $k$-th cluster, as follows:
for $k=1,2,\cdots$,
\begin{align*}
\bar {a} \langle k \rangle &:= \begin{bmatrix} \bar a[(k-1)\beta] \\ \bar a[(k-1)\beta+1] \\ \vdots \\ \bar a[k\beta-1] \end{bmatrix} \in \mathbb{R}^{\beta p},  \\
\tilde x \langle k \rangle &:= \begin{bmatrix} \tilde x(((k-1)\alpha+1_\delta)T_s) \\ \tilde x(((k-1)\alpha+2_\delta)T_s) \\ \vdots \\ \tilde x(((k-1)\alpha+\alpha_\delta)T_s) \end{bmatrix}\in\mathbb{R}^{\alpha n}, \\
\tilde{y} \langle k \rangle &:= {\mathcal C} \tilde x \langle k \rangle \in \mathbb{R}^{\alpha q} 
\end{align*}
where ${\mathcal C} := I_\alpha \otimes C$ ($I_\alpha$ is the identity matrix of size $\alpha$ and $\otimes$ is the Kronecker product).
It is noted that the vector $\tilde y \langle k \rangle$ is the collection of $\alpha$ measurements within one cluster.

Now, let us focus on the terminal state of each cluster, which is denoted by $\tilde x_\ccc [ k ] := \tilde x(k\beta T_a) \in \mathbb{R}^n$.
Then, from \eqref{xa dynamics}, one can derive that
\begin{align}\label{terminal_state}
\begin{split}
\tilde x_\ccc [k] &= A_\ddd^\beta\tilde x_\ccc [k-1] \\
&\qquad + \begin{bmatrix} A_\ddd^{\beta-1}B_\ddd & A_\ddd^{\beta-2}B_\ddd & \cdots & B_\ddd \end{bmatrix} \bar a\langle k \rangle \\
&=: A_\ddd^\beta \tilde x_\ccc [k-1] + \Phi_\ccc \bar a  \langle k \rangle 
\end{split}
\end{align}
with $\tilde x_\ccc [0] = 0$.
Similarly, one can derive the following for $\tilde x \langle k \rangle$.

\begin{lem}\label{lem_Pi}
It follows that
\begin{align}
\tilde x \langle k \rangle &= \bar A_\alpha \tilde x_\ccc[k-1] + \Pi \bar a \langle k \rangle \label{eq:Ts1} \\
\tilde y \langle k \rangle &= {\mathcal C} \bar A_\alpha \tilde x_\ccc[k-1] + {\mathcal C} \Pi \bar a \langle k \rangle \label{eq:Ts2}
\end{align}
where
\begin{align*}
\bar A_\alpha := \begin{bmatrix} e^{A 1_\delta T_s} \\ e^{A 2_\delta T_s} \\ \vdots \\ e^{A \alpha_\delta T_s} \end{bmatrix}
\end{align*}
and the $(l,m)$-th block of $\Pi \in \mathbb{R}^{\alpha n \times \beta p}$ is defined as follows: 
\begin{align}\label{N>1Pi}
\Pi{(l,m)}:=\begin{cases}
A_\ddd^{\langle l_\delta , m \rangle} B_\ddd , &m=1,\cdots,\lfloor l_\delta R 	\rfloor , \\
B_\ddd^{\langle l_\delta , \lfloor l_\delta R \rfloor \rangle } , &m=\lfloor l_\delta R \rfloor +1,\\ 
0, &m=\lfloor l_\delta R \rfloor +2 , \cdots ,\beta,
\end{cases}
\end{align}
for $l=1,\cdots,\alpha.$ 
\end{lem}

{\it Proof:}
Consider the first cluster $k=1$, in which the state $\tilde x(0)$ at the beginning of the cluster is zero and $\tilde x_\ccc[0]=0$.
With the property $\lfloor j_\delta R \rfloor T_a \le j_\delta R T_a = j_\delta T_s$, one can compute the state $\tilde x(j_\delta T_s)$, whose sensing time $j_\delta T_s$ belongs to this cluster, by the variation of constant formula as follows:
\begin{align}\label{derivation_Pi}
\begin{split}
&\tilde x(j_\delta T_s) = \int_0^{T_a} e^{A(j_\delta T_s - \tau)} d\tau B \bar a[0] \\
&\quad + \int_{T_a}^{2T_a} e^{A(j_\delta T_s - \tau)} d\tau B \bar a[1] \; + \; \cdots \\
&\quad + \int_{(\lfloor j_\delta R \rfloor - 1)T_a}^{(\lfloor j_\delta R \rfloor)T_a} e^{A(j_\delta T_s - \tau)} d\tau B \bar a[\lfloor j_\delta R \rfloor - 1] \\
&\quad + \int_{\lfloor j_\delta R \rfloor T_a}^{j_\delta T_s} e^{A(j_\delta T_s - \tau)} d\tau B \bar a[\lfloor j_\delta R \rfloor] \\
&= \sum_{m=1}^{\lfloor j_\delta R \rfloor} e^{A(j_\delta T_s - m T_a)} \int_{(m-1)T_a}^{mT_a} e^{A(mT_a-\tau)} d\tau B \bar a[m-1] \\
&\quad + \int_{0}^{j_\delta T_s - \lfloor j_\delta R \rfloor T_a} e^{A\tau} d\tau B \bar a[\lfloor j_\delta R \rfloor] \\
&= \sum_{m=1}^{\lfloor j_\delta R \rfloor} A_\ddd^{\langle j_\delta,m\rangle} B_\ddd \bar a[m-1] 
+ B_\ddd^{\langle j_\delta, \lfloor j_\delta R \rfloor  \rangle} \bar a[\lfloor j_\delta R \rfloor] .
\end{split}
\end{align}
When $\lfloor 1_\delta R \rfloor = 0$ (which happens if $j=1$ and $1_\delta T_s < T_a$), it should be interpreted that the summation term in the above equation is zero or null.
The discussion so far verifies \eqref{eq:Ts1} and the matrix $\Pi$ for $k=1$.

For the general $k$-th clusters ($k > 1$), the derivation is the same (because the pattern for actuation and sensing times are repeated along the clusters) except that the state $\tilde x_\ccc[k-1] = \tilde x((k-1)\beta T_a)$ need not be zero.
Taking into account $\tilde x_\ccc[k-1]$ as the initial condition for the corresponding cluster, one can easily verify \eqref{eq:Ts1} for $k > 1$.
Once \eqref{eq:Ts1} is verified, \eqref{eq:Ts2} trivially follows. $\hfill\blacksquare$

Now, let us define the {\it disruption time} 
$$t_k:= (k-1)\beta T_a + T_k^*, \qquad T_k^* \in (0, \beta T_a],$$
which is the time when the disruptive property is met within the $k$-th cluster.
The sequence $\{T_k^*\}_{i=1}^\infty$ is chosen by adversary, and it is often a fixed number, for convenience, like $T_k^* = \beta T_a$ or $T_k^* = \beta T_a/2$ for all $k$.
For simplicity of presentation, let us normalize the disruption time as $t_k^*:= T_k^*/(\beta T_a) \in (0,1]$.
Then, the error state at the disruption time $t_k$, which we will denote as $\tilde x_{a} [ k ] := \tilde x (t_k)$, is computed as follows.

\begin{lem}\label{lem_Phi}
It follows that
\begin{align}\label{disruptive_time}
\tilde x_a[k] = \bar A_k^* \tilde x_\ccc [k-1] + \Phi_k^* \bar a \langle k \rangle
\end{align}
where $\bar A_k^* := e^{AT_k^*}=e^{At_k^*\beta T_a}$ and
	\begin{align}\label{Phi_form}
	\Phi_k^* &= [\Phi_k^*(1,1), \cdots, \Phi_k^*(1,\beta)] \in \mathbb{R}^{n \times \beta p} \\
	\Phi_k^*(1,m) &= 
	\begin{cases}
	A_\ddd^{\langle 0, m - \beta t_k^* \rangle} B_\ddd, &m = 1,\cdots, \lfloor \beta t_k^* \rfloor \\
	B_\ddd^{\langle 0, \lfloor \beta t_k^* \rfloor - \beta t_k^* \rangle}, &m= \lfloor \beta t_k^* \rfloor + 1\\
	0, &m= \lfloor \beta t_k^*   \rfloor +2 , \cdots, \beta. \notag
	\end{cases}
	\end{align}
\end{lem}

{\it Proof:}
The proof is similarly done as Lemma \ref{lem_Pi}.
For the first cluster ($k=1$), the state $\tilde x_a$ at time $t_1$ is evaluated similarly as \eqref{derivation_Pi} with $j_\delta T_s$ being replaced by $t_1 = \beta t_1^* T_a$, and $j_\delta R$ being replaced by $\beta t_1^*$. 
Indeed, it follows that
\begin{align*}
\begin{split}
&\tilde x_a [1]=\tilde x (t_1) = \tilde x (T_1^*) \\
&= \sum_{m=1}^{\lfloor \beta t_1^* \rfloor} e^{A(\beta t_1^* T_a - m T_a)} \int_{(m-1)T_a}^{mT_a}e^{A(mT_a -\tau )}d\tau B \bar a [m-1]\\ 
&\quad + \int_{0}^{\beta t_1^* T_a - \lfloor \beta t_1^* \rfloor T_a } e^{A\tau}d\tau B \bar a [\lfloor \beta t_1^* \rfloor ]\\
&= \sum_{m=1}^{\lfloor \beta t_1^* \rfloor}A_\ddd^{\langle 0,m - \beta t_1^*  \rangle} B_\ddd \bar a [m-1] + B_\ddd^{\langle 0, \lfloor  \beta t_1^* \rfloor - \beta t_1^* \rangle} \bar a [\lfloor \beta t_1^* \rfloor].
\end{split}
\end{align*}
Like in Lemma \ref{lem_Pi}, if $\lfloor \beta t_1^* \rfloor = 0$, the summation term in the above equation becomes zero.
Thus, \eqref{disruptive_time} and \eqref{Phi_form} are verified for the first cluster.

For the case $k>1$, by taking into account the initial condition $\tilde x_\ccc[k-1]$ and by noting that the matrix $\Phi_k^*$ is obtained exactly the same way as for $k=1$, equation \eqref{disruptive_time} is easily verified.
$\hfill\blacksquare$

Note that, if all $T_k^*$ are chosen as a constant for all $k \ge 1$, then both $\bar A_k^*$ and $\Phi_k^*$ are constant matrices.
Now, with Lemma \ref{lem_Pi} and Lemma \ref{lem_Phi}, the problem of our interest is reformulated in a cluster-wise sense; i.e., our interest becomes designing an attack sequence $\bar a \langle k \rangle $ that satisfies $\|\tilde x_a [k]\|\ge H_k $ (disruptive property), and at the same time, $\tilde y \langle k \rangle \equiv 0$ for each $k$-th cluster (zero-stealthy property) for all $k \ge 1$.

\subsection{Conditions for Attack Design}

With equations \eqref{eq:Ts1}, \eqref{eq:Ts2}, and \eqref{disruptive_time} at hand, conditions for attack design can be established.
First of all, by \eqref{eq:Ts2}, stealthiness of the attack is obtained if the attack sequence $\bar a \langle k \rangle$ for the $k$-th cluster belongs to the kernel of ${\mathcal C} \Pi$, and so, we require the kernel is non-trivial.
Second, for the disruptive property of the state $\tilde x_a[k]$ in \eqref{disruptive_time}, we ask the kernel of $\Phi_k^*$ not to include the kernel of ${\mathcal C}\Pi$ because, if $\ker \Phi_k^* \supset \ker \mathcal{C}\Pi$, then any stealthy attack has no affect on $\tilde x_a[k]$.
Finally, as the attack is initiated, the state $\tilde x(t)$ becomes non-zero, and therefore, even if the attack $\bar a \langle k-1 \rangle$ is designed to be stealthy from the measurement vector $\tilde y \langle k-1 \rangle$ for the $(k-1)$-th cluster, it may become detectable through non-zero $\tilde x_\ccc[k-1] = \tilde x((k-1)\beta T_a)$ in the $k$-th cluster. 
See \eqref{eq:Ts1} and \eqref{eq:Ts2}.
In order to counteract it, we require the range space of ${\mathcal C}\bar A_\alpha$ would belong to the range space of ${\mathcal C}\Pi$ so that some component of the attack sequence $\bar a \langle k \rangle$ is designed to cancel the effect of $\tilde x_\ccc[k-1]$ on $\tilde y \langle k \rangle$. 
These discussions yield the following formal assumption.

\begin{asm}\label{asm: input redundancy}
The following conditions hold:
\begin{itemize}
\item[(a)] $\ker \mathcal{C}\Pi \neq \{0\}$,
\item[(b)] $\ker \mathcal{C}\Pi \not \subset \ker \Phi_k^*$, $k \ge 1$, with disruption times $\{t_k^*\}_{i=1}^\infty$,
\item[(c)] ${\rm im}\; \mathcal{C} \bar A_\alpha \subset {\rm im}\; \mathcal{C}\Pi$. \hfill$\square$
\end{itemize}
\end{asm}

A few sufficient conditions for Assumption \ref{asm: input redundancy} can be derived.
For example, since $\mathcal {C}\Pi \in \mathbb{R}^{\alpha q\times \beta p}$ so that $\alpha q < \beta p$ implies $\ker \mathcal{C}\Pi \neq \{0\}$, the item (a) is satisfied either when the number $p$ of inputs is large, or when the actuator works faster than the sensor (i.e., $R = T_s/T_a = \beta/\alpha$ is large enough). 
Hence, a sufficient condition for the item (a) is obviously $qT_a < pT_s$, which is simpler to check than item (a).
On the other hand, it is noted that the condition (c) holds if the matrix $\mathcal{C}\Pi$ has full row rank or if the matrix $\Pi$ has full row rank.
Finally, for the condition (b), we have the following.

\begin{prop}\label{prop:b}
	If the condition (a) of Assumption \ref{asm: input redundancy} holds and $B_\ddd$ has full column rank (i.e., ${\rm rank}\; B_\ddd = p$), then there exists a sequence $\{t_k^*\}_{i=1}^\infty$ with which the condition (b) holds.
\end{prop}

{\it Proof}: 
By the condition (a), pick any non-zero $z = {\rm col}(z_1,\cdots,z_\beta) \in \ker \mathcal{C}\Pi$ where $z_i \in \R^p$.
Define the index $i^* := \min \{i: z_i \neq 0, i=1,\dots,\beta\}$, and pick the disruption time $t_k^* \in (0,1]$ such that $i^*= \beta t_k^*$.
Then, it follows from \eqref{Phi_form} that $\Phi_k^* z = B_\ddd z_{i^*} \neq 0$ since $B_\ddd$ has full column rank.
This implies that $\ker \mathcal{C}\Pi \not\subset \ker \Phi_k^*$, i.e., the item (b).
\hfill$\blacksquare$

\begin{rem}\label{N1}
As a special case, let us consider the case when $R$ is a positive integer (i.e., $R = N \ge 1$ so that $\alpha=1$ and $\beta=N$), and $\delta = 0$.
This is the case that has been studied in \cite{KPSE16}.
In this case, we have $\mathcal{C} = C$, $\bar A_\alpha = e^{AT_s} = A_\ddd^N$, and $\Pi = [e^{A(T_s-T_a)}B_\ddd, e^{A(T_s-2T_a)}B_\ddd, \cdots, e^{A(T_s-(N-1)T_a)}B_\ddd, B_\ddd] = [A_\ddd^{N-1}B_\ddd,A_\ddd^{N-2}B_\ddd\cdots,B_\ddd]$, and the conditions (a) and (c) of Assumption \ref{asm: input redundancy} read as
\begin{itemize}
\item [(a)] $\{0\} \neq \ker C[A_\ddd^{N-1}B_\ddd,\cdots,B_\ddd]$, 
\item [(c)] ${\rm im}~CA_\ddd^N \subset {\rm im}~C[A_\ddd^{N-1}B_\ddd,\cdots,B_\ddd]$.
\end{itemize}
It is clear that the above conditions hold if $q < Np$ and if either $C[A_\ddd^{N-1}B_\ddd,\cdots,B_\ddd]$ or $[A_\ddd^{N-1}B_\ddd,\cdots,B_\ddd]$ has full row rank.
On the other hand, in \cite{KPSE16}, the disruption time $t_k^*$ is determined in the assumption as one of $\{1/N, 2/N, \cdots, 1\}$.
It is noted from \eqref{Phi_form} that, for each candidate of $t_k^* = j/N$, $j=1,\cdots,N$, the matrix $\Phi_k^*$ becomes as $\Phi_k^*|_{t_k^*=j/N} = [A_\ddd^{j-1}B_\ddd, A_\ddd^{j-2}B_\ddd, \cdots, B_\ddd, 0, \cdots, 0]$.
To facilitate selection of $t_k^*$ among the candidates, the condition of \cite{KPSE16} reads as 
\begin{multline*}
\text{(b')} \;\; \ker C[A_\ddd^{N-1}B_\ddd,\cdots,B_\ddd] \\
\not \subset \ker 
\begin{bmatrix} B_\ddd & 0 & \cdots & 0 \\
\vdots & \vdots & \ddots & \vdots \\
A_\ddd^{N-2}B_\ddd & A_\ddd^{N-3}B_\ddd & \cdots & 0 \\
A_\ddd^{N-1}B_\ddd & A_\ddd^{N-2}B_\ddd & \cdots & B_\ddd \end{bmatrix}.
\end{multline*}
When this condition holds, one can pick suitable $t_k^*$ among the candidates for the condition (b) of Assumption \ref{asm: input redundancy}.
Another sufficient condition for (b) in this special case is: (b'') $\ker C \cap \;{\rm im}~\Pi \not = \{0\}$.
This is because (b'') means that there exists a vector $v$ such that $\Pi v \not = 0$ and $\Pi v \in \ker C$. 
This implies that the vector $v$ belongs to $\ker C\Pi$ while it does not belong to $\ker \Pi$, which guarantees (b') with $t_k^*=1$.
In Section \ref{sec:N=1example}, we demonstrate this case with $N=1$. \hfill $\square$
\end{rem}

\subsection{Off-line Construction of Attack Signal}

\begin{figure}[t!]
\begin{center}
\includegraphics[width=0.38\textwidth]{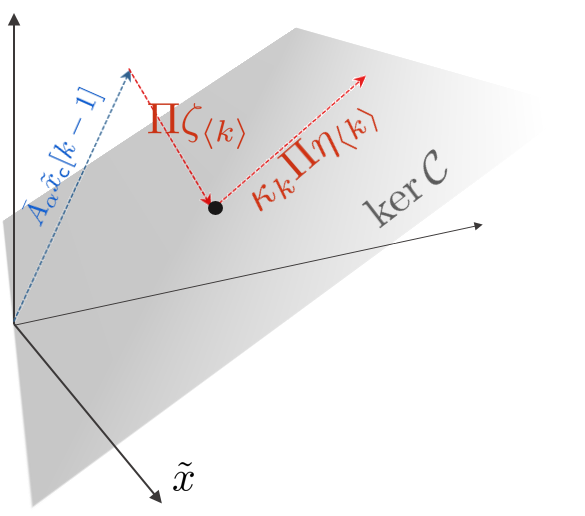}
\end{center}
\caption{Graphical interpretation of attack components}\label{fig:attackconcept}
\end{figure}

In this subsection, based on Assumption \ref{asm: input redundancy}, we design an attack sequence $\bar a[i]$, or equivalently $\bar a \langle k \rangle$, that solves the reformulated problem; i.e., to make $\|\tilde x_a[k]\| \ge H_k$ and $\tilde y \langle k \rangle \equiv 0$ for $k=1,2,\cdots$. 
In particular, we propose the sequence $\bar a \langle k \rangle$ in the following form:
$$\bar a \langle k \rangle = \kappa_k \eta_{\langle k \rangle} + \zeta_{\langle k \rangle} \in \R^{\beta p}$$
where $\kappa_k$ is a positive constant and $\eta_{\langle k \rangle}, \zeta_{\langle k \rangle} \in \R^{\beta p}$.
The idea is to pick $\eta_{\langle k \rangle}$ such that $\Pi \eta_{\langle k \rangle}$ is stealthy (i.e., belongs to $\ker \mathcal{C}$) but disruptive (i.e., $\Phi_k^*\eta_{\langle k \rangle} \not = 0$) while $\kappa_k$ decides the intensity of disruption, and to pick $\zeta_{\langle k \rangle}$ to counteract the effect of non-zero $\tilde x_\ccc[k-1]$ on $\tilde y \langle k \rangle$ (i.e., $\bar A_\alpha \tilde x_\ccc[k-1] +\Pi \zeta_{\langle k \rangle} \in \ker \mathcal{C}$).
See Fig.~\ref{fig:attackconcept}.

The attack signal is designed sequentially, i.e., in the order of $\bar a \langle 1 \rangle$, $\bar a \langle 2 \rangle$, and so on.
As the first step, let $\zeta_{\langle 1 \rangle} = 0$ (since there is no attack before the time $t=0$), and pick $\eta_{\langle 1 \rangle} \in \ker \mathcal{C}\Pi$ such that $\Phi_1^* \eta_{\langle 1 \rangle} \not = 0$ (whose existence is guaranteed by Assumption \ref{asm: input redundancy}.(a)).
Then, stealthiness follows since
\begin{align}\label{ya=0}
\begin{split}
\tilde{y} \langle 1 \rangle  = \mathcal{C}\Pi \bar{a}\langle 1 \rangle = \kappa_1 \mathcal{C}\Pi \eta_{\langle 1 \rangle} = 0.
\end{split}
\end{align}
For the disruptive property, pick $\kappa_1>0$ such that
\begin{align}
\kappa_1 \|\Phi_1^* \eta_{\langle 1 \rangle}\| \ge H_1 \label{M1}
\end{align}
By this, a stealthy and disruptive attack signal $\bar a[0]$, $\cdots$, $\bar a[\beta-1]$ is obtained for the first cluster $(0,\beta T_a]$.

In order to design $\bar a \langle 2 \rangle$ for the second cluster, consider 
$$\tilde{y} \langle 2 \rangle =\mathcal{C} \bar{A}_\alpha \tilde x_\ccc[1]  +\mathcal{C}\Pi \bar{a} \langle 2 \rangle$$
where $\tilde x_\ccc[1]$ is computed by \eqref{terminal_state} and $\bar a \langle 2 \rangle = \kappa_2 \eta_{\langle 2 \rangle} + \zeta_{\langle 2 \rangle}$.
Similarly as before, pick $\eta_{\langle 2 \rangle}$ such that
\begin{equation}\label{eq:alpha2}
\mathcal{C}\Pi \eta_{\langle 2 \rangle} = 0 \quad \text{and} \quad \Phi_2^* \eta_{\langle 2 \rangle} \not = 0
\end{equation}
and pick $\zeta_{\langle 2 \rangle}$ such that
\begin{equation} 
\mathcal{C} \Pi \zeta_{\langle 2 \rangle} = -  \mathcal{C} \bar A_\alpha \tilde x_\ccc[1]. \label{beta2}
\end{equation}
Since 
\begin{align}
\tilde x_a [2] &=
 \bar {A}_2^* \tilde x_\ccc[1]
+ \Phi_2^* \bar a \langle 2 \rangle \notag \\
&= \bar {A}_2^* \tilde x_\ccc[1] + \Phi_2^* \zeta_{\langle 2 \rangle} + \Phi_2^* \kappa_2 \eta_{\langle 2 \rangle} ,\label{xtilde}
\end{align}
we take $\kappa_2$ such that
\begin{equation}\label{ka2}
\kappa_2 \| \Phi_2^* \eta_{\langle 2 \rangle} \| \ge H_2 + \| \bar A_2^* \tilde x_\ccc[1] + \Phi_2^* \zeta_{\langle 2 \rangle} \|.
\end{equation}
This ensures stealthiness and disruptive property of the attack in the second cluster $(\beta T_a , 2\beta T_a ]$.

We now generalize the procedure.

\noindent{\bf Procedure of Attack Signal Generation}:
\nopagebreak

\noindent{\it Step }$k$ ($k = 1, 2 \dots $): Take $\zeta_{\langle k \rangle}$ so that the following equation holds: 
\begin{align}\label{beta}
\mathcal{C}\Pi \zeta_{\langle k \rangle} = -\mathcal{C}\bar A_\alpha \tilde{x}_\ccc[k-1]
\end{align}
(for $k=1$, $\tilde x_\ccc[0] = 0$ so that $\zeta_{\langle 1 \rangle}=0$).
Pick $\eta_{\langle k \rangle}$ such that 
\begin{equation*}
\eta_{\langle k \rangle} \in \ker\mathcal{C}\Pi \quad \text{and} \quad \eta_{\langle k \rangle} \notin \ker \Phi_k^*
\end{equation*}
and select a positive $\kappa_k$ such that
\begin{align}\label{kappa3}
\kappa_k \ge
\dfrac{ H_k +  \| \bar A_k^* \tilde x_\ccc[k-1] + \Phi_k^* \zeta_{\langle k \rangle} \| }{\| \Phi_k^* \eta_{\langle k \rangle}\| }.
\end{align}
With these terms, construct $\bar{a} \langle k \rangle = \kappa_k \eta_{\langle k \rangle} + \zeta_{\langle k \rangle}$ and update $\tilde x_\ccc[k] = A_\ddd^\beta\tilde x_\ccc[k-1] + \Phi_\ccc \bar a\langle k \rangle$ by \eqref{terminal_state}. \hfill$\blacksquare$

\begin{rem}\label{alpha}
It is noted that the construction of attack sequence can be done off-line, or {\it a priori} before the attack begins, because the procedure does not need any real-time information.
Moreover, if the normalized disruption time $t_k^* \in (0,1]$ is chosen as a fixed constant for all $k \ge 1$, then the matrices $\Phi_k^*$ are the same for all $k \ge 1$.
Then, the vector $\eta_{\langle k \rangle}$ can also be chosen as a constant $\eta$.
\hfill$\square$
\end{rem}

\medskip

We close this section by summarizing the discussions so far.
\begin{thm}\label{main theorem}
Suppose that the adversary has the information of $T_s$, $T_a$, and $\Delta$ as well as the system information of $A$, $B$, and $C$.
If $R=T_s/T_a \in \mathbb{Q}$ and Assumption \ref{asm: input redundancy} holds with normalized disruption times $\{t_k^*\}_{k=1}^\infty$, then an attack sequence $\{\bar a[i]\}_{i=0}^{\infty}$ constructed via the proposed procedure has the zero-stealthy property and the disruptive property for any given $\{H_k\}_{k=1}^{\infty}$. $\hfill\square$
\end{thm}

\section{Examples}\label{sec:Sim} 

\subsection{Numerical Example: $R=1$ with $\delta=0$}\label{sec:N=1example}

In this subsection, we study a simple example in order to illuminate the attack generation procedure for the case $R=1$ without offset, as discussed in Remark \ref{N1}.
For this, let us consider the error dynamics \eqref{xasys} with
\begin{equation}
A = \begin{bmatrix}
-1 &0 &0 \\0 &-5 &-3 \\ 0 &2 &0
\end{bmatrix},~~B=\begin{bmatrix}
1 & 0 \\ 0 & 1 \\ 0 & 0
\end{bmatrix},~~C=\begin{bmatrix}
1 & 0 & 1
\end{bmatrix}.\notag
\end{equation}
With a zero-order holder and sampler whose sampling periods are $T_a=T_s=1~{\rm sec}$ (and thus $R=1$), its sampled-data system is given by \eqref{xa dynamics} with
$$A_\ddd = \begin{bmatrix}
0.368 & 0 & 0 \\ 0 & -0.121 & -0.257 \\ 0 & 0.171 & 0.306 \end{bmatrix},
\;\;
B_\ddd = \begin{bmatrix}
0.632 & 0 \\ 0 & 0.086 \\ 0 & 0.231
\end{bmatrix}.$$
It is easily seen that Assumption \ref{asm: input redundancy} holds for the above system (see Remark \ref{N1}).
In particular, $q<Rp$ so that (a) holds and the matrix $\mathcal{C}\Pi = CB_\ddd$ has full row rank so that (c) holds.
Also, the matrix $B_\ddd$ has full column rank, and so, by (the proof of) Proposition \ref{prop:b}, the condition (b) holds with $t_k^*=i^*/\beta=1/1$.

Now, for given $H_k=k$, an attack sequence $\bar a \langle k \rangle = \kappa_{\langle k \rangle} \eta_{\langle k \rangle} + \zeta_{\langle k \rangle}$ is constructed as follows: 

{\it Step 1}: Set $\zeta_{\langle 1 \rangle}=\rm{col}(0,0)$ and $\eta_{\langle 1 \rangle}= \rm{col}(-0.343,0.939) \in \ker \mathcal{C}\Pi$ such that $\Phi_k^* \eta_{\langle 1 \rangle} = B_\ddd \eta_{\langle 1 \rangle} \not = 0$. 
Then, select $\kappa_1=3.15$ to satisfy \eqref{M1} (i.e., $\kappa_1 \|B_\ddd \eta_{\langle 1 \rangle}\| \ge H_1$). 
Set $\tilde{x}_\ccc[1] = B_\ddd(\kappa_1\eta_{\langle 1 \rangle}+\zeta_{\langle 1 \rangle})$.

{\it Step 2}: Choose $\zeta_{\langle 2 \rangle}$ such that \eqref{beta2} holds (i.e., $CB_\ddd \zeta_{\langle 2 \rangle}=-CA_\ddd\tilde{x}_\ccc[1]$). 
For convenience, let $\eta_{\langle 2 \rangle} = \eta_{\langle 1 \rangle}$ as discussed in Remark \ref{alpha}.  
Then, select $\kappa_2$ for \eqref{ka2}, and set $\tilde x_\ccc [ 2 ] = A_\ddd \tilde{x}_\ccc [ 1 ] + B_\ddd(\kappa_2\eta_{\langle 2 \rangle}+\zeta_{\langle 2 \rangle})$. 

Similarly, the remaining steps proceed with  $\eta_{\langle k \rangle}=\eta_{\langle 1 \rangle}$.

\begin{figure}[t!]
	\begin{center}
		\includegraphics[width=0.49\textwidth]{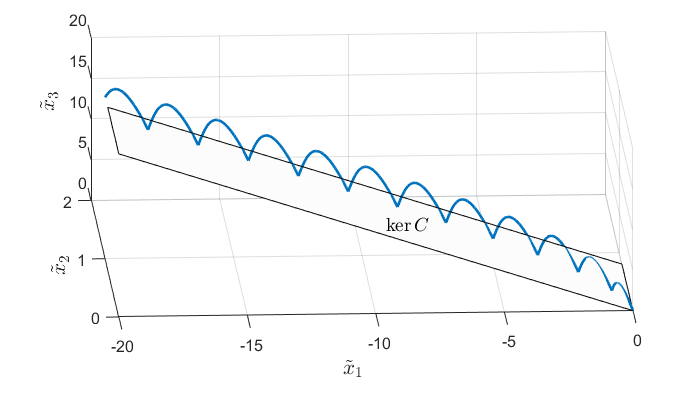}
	\end{center}
	\caption{Continuous-time state trajectory $\tilde{x}(t)$ (solid blue line) and $\ker C$ (plane)}\label{fig:N1}
\end{figure}

\begin{figure}[t!]
	\begin{center}
		\includegraphics[width=0.5\textwidth]{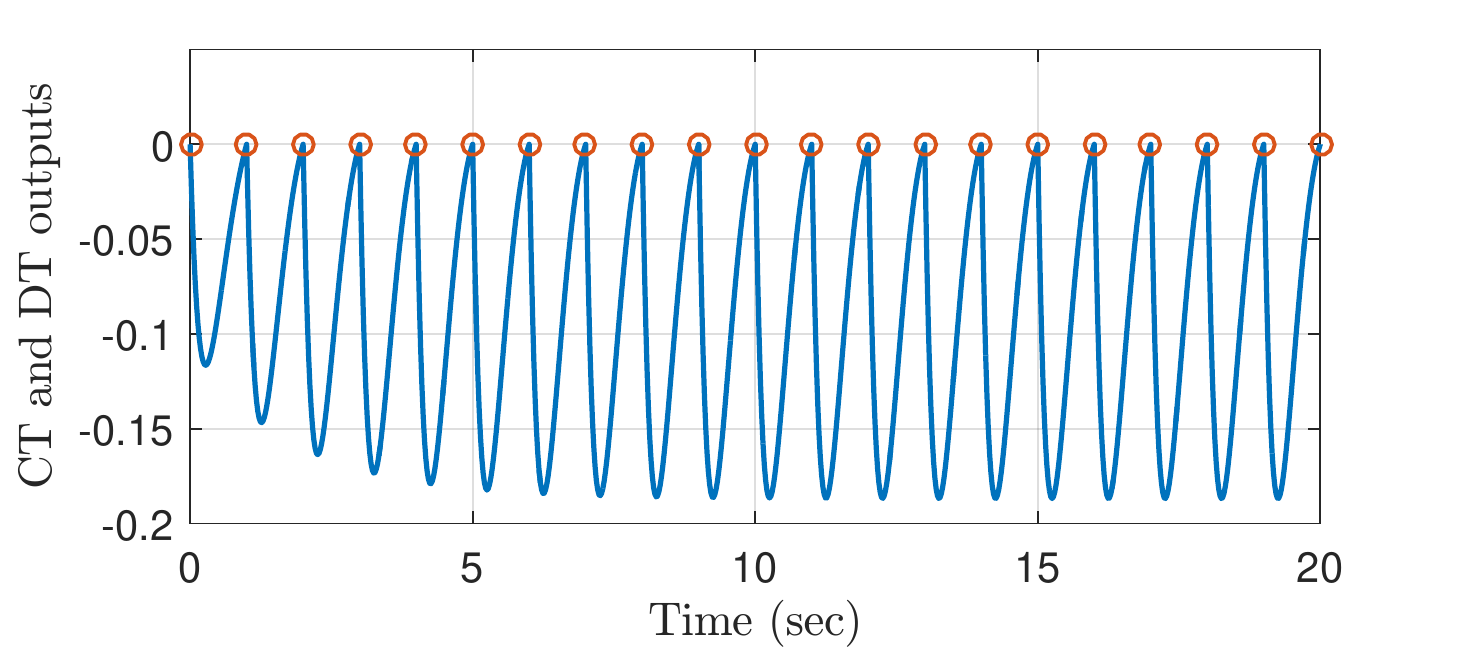}
	\end{center}
	\caption{Continuous-time output $\tilde{y}(t)$ (solid blue line) and discrete-time output $\tilde{y}(jT_s)$ (red circle)}\label{fig:N1Y}
\end{figure}

\begin{figure}[t!]
	\begin{center}
		\includegraphics[width=0.5\textwidth]{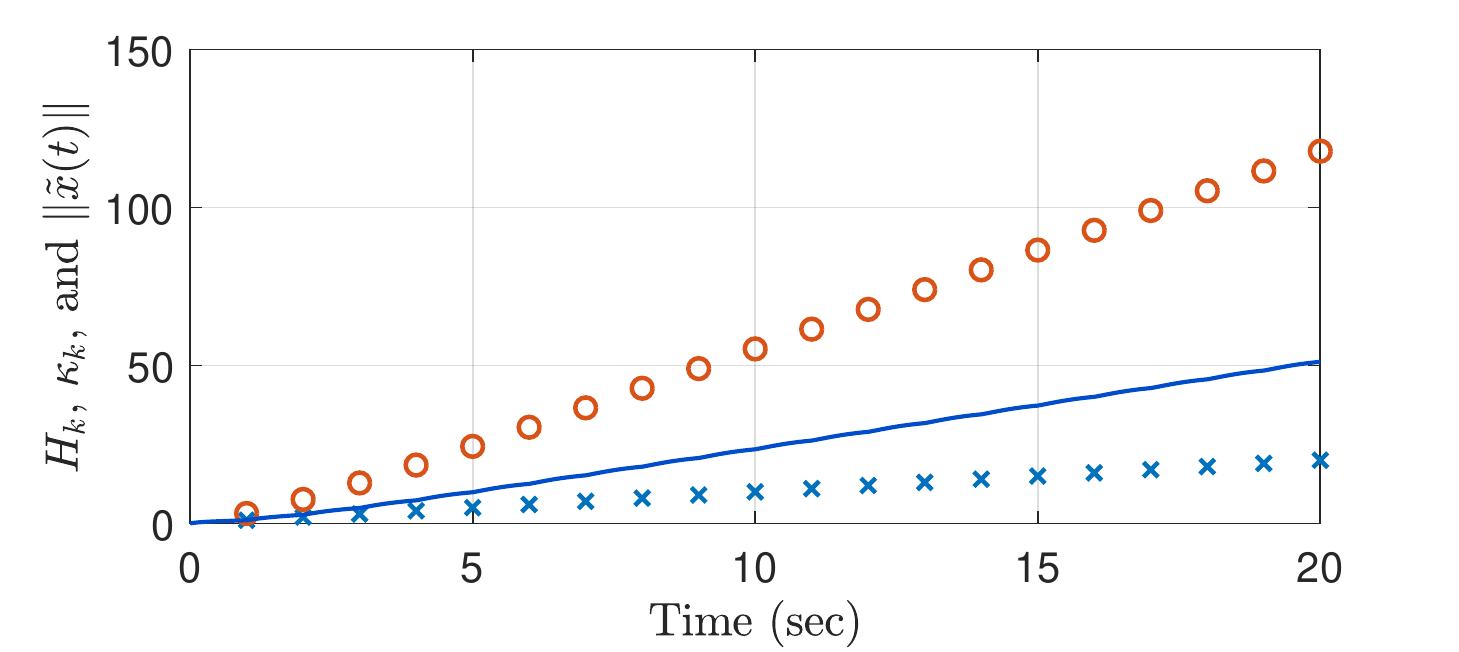}
	\end{center}
	\caption{Sequence $H_k$ (blue cross), selected $\kappa_k$ (red circle), and $\|\tilde{x}(t)\|$ (blue solid line)}\label{fig:Mkappa}
\end{figure}

The designed attack sequence is injected into the control input at $t=0~{\rm sec}$. 
Fig.~\ref{fig:N1} shows the continuous-time state $\tilde{x}(t)$ from its initial condition $\tilde{x}(0)={\rm col}(0,0,0)$.
Note that $\tilde x(t)$ is the error between the attack-free state $x_{\rm o}(t)$ and the state $x(t)$ under attack.
In this figure, it is observed that the error $\tilde x(t)$ moves far from the origin while it repeatedly encounters $\ker C$. 
The sampled error output $\tilde y(jT_s)$ remains zero as seen in Fig.~\ref{fig:N1Y}) ({\it zero-stealthy} property). 
On the other hand, from Fig.~\ref{fig:Mkappa}, one can see that the {\it disruptive property} is satisfied; that is, $\|\tilde{x}(t_k)\|\ge H_k$ for $(k-1)T_a <  t_k \leq kT_a$ (here, $t_k = kT_a$).

\subsection{X-38 Vehicle Example: $R=4$ with $\delta=0$}\label{sec: simulation result X-38} 

As another example, we consider the X-38 vehicle model which is a prototype flight test vehicles for crew return vehicle \cite{LW00}. 
In \cite{LW00}, the X-38 is operated by a multi-rate digital controller whose holder operates four times faster than the sampler (i.e., $R=T_s/T_a=4$) with $T_a=0.04~{\rm sec}$ and $T_s=0.16~{\rm sec}$.
The X-38 model has $3$ inputs, $9$ outputs, and $11$ states ($A\in\mathbb{R}^{11\times11}$, $B\in\mathbb{R}^{11\times3}$, and $C\in\mathbb{R}^{9\times11}$). 
More detailed information on the X-38 plant is provided in \cite{LW00}, \cite{JJ98}.

From the information of X-38 model in \cite{LW00} (that is omitted in this paper), one can verify that Assumption \ref{asm: input redundancy} holds by the following reasons:
\begin{itemize}
	\item $Rp=12$ and $q=9$, and so, the condition (a) holds (i.e., $\mathcal{C}\Pi \in \R^{9 \times 12}$ so that $\ker \mathcal{C}\Pi \neq \{0\}$),
	\item the matrix $B_\ddd$ has full column rank, and there exists a non-zero vector $z$ such that $\mathcal{C}\Pi z = 0$ where the first 3 components are a non-zero vector in $\R^3$. Then, by the proof of Proposition \ref{prop:b}, $i^*=1$. Therefore, the condition (b) holds with $t_k^* = i^*/\beta = 1/4$,
	\item the matrix $\mathcal{C}\Pi$ has full row rank so that ${\rm im}~\mathcal{C}\Pi=\mathbb{R}^9$ and the condition (c) holds.
\end{itemize}

Now, following the proposed attack generation procedure, we construct an attack sequence $\bar{a} \langle k \rangle = \kappa_k \zeta_{\langle k \rangle} + \eta_{\langle k \rangle}$ with disruptive property $H_k = 0.5k$. 
In particular, we have chosen
\begin{align*}
\eta_k &= \eta \\
&= \text{col}\left(\begin{bmatrix}
-0.132\\0.145\\0.108
\end{bmatrix},\begin{bmatrix}
0.397\\-0.434\\-0.324
\end{bmatrix},\begin{bmatrix}
-0.396\\0.434\\0.324
\end{bmatrix},\begin{bmatrix}
0.132\\-0.144\\-0.108
\end{bmatrix}\right) \\ 
&\in \ker \mathcal{C}\Pi \qquad \text{and} \qquad \Phi_k^* \eta \not = 0,
\end{align*} 
and $\zeta_{\langle k \rangle}$'s and $\kappa_k$'s are selected to satisfy \eqref{beta} and \eqref{kappa3}, respectively.

\begin{figure}[t!]
	\begin{center}
		\includegraphics[width=0.5\textwidth]{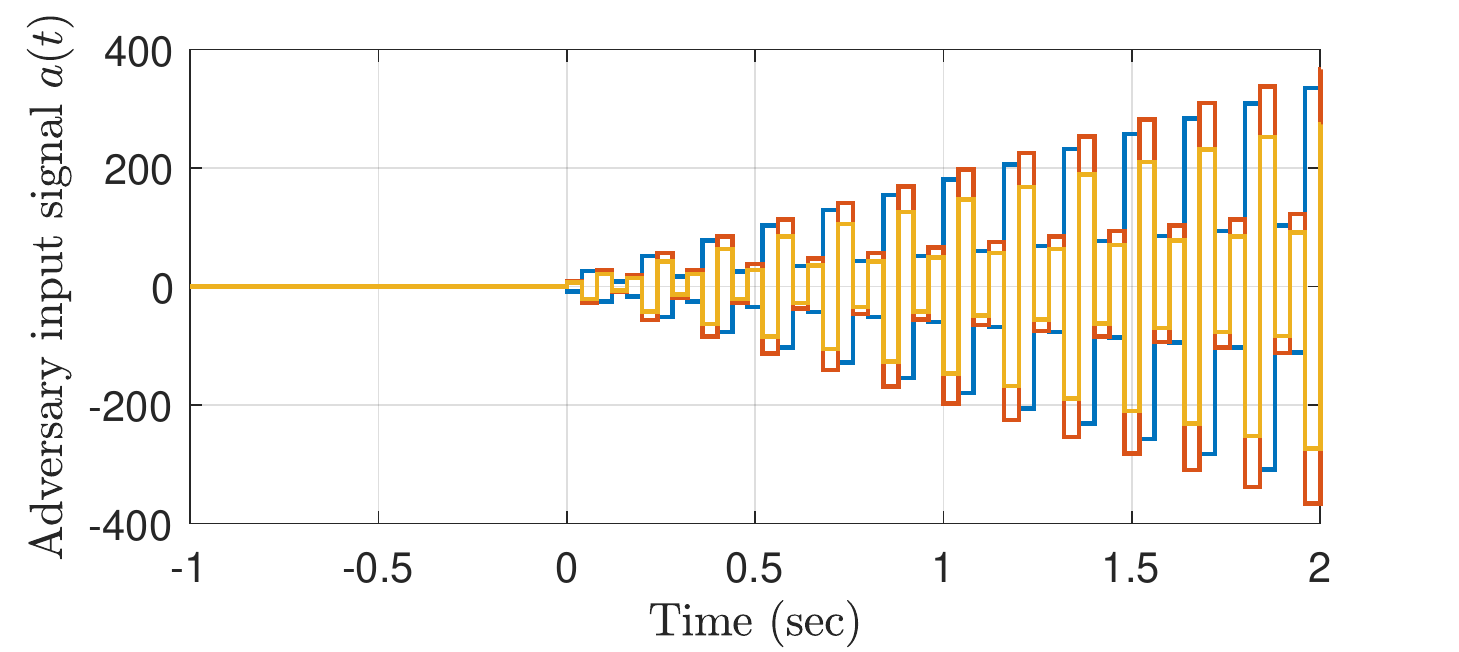}
	\end{center}
	\caption{Zero-stealthy attack signal $a(t) $}\label{fig: attack}
\end{figure}

\begin{figure}[t!]
	\begin{center}
		\includegraphics[width=0.5\textwidth]{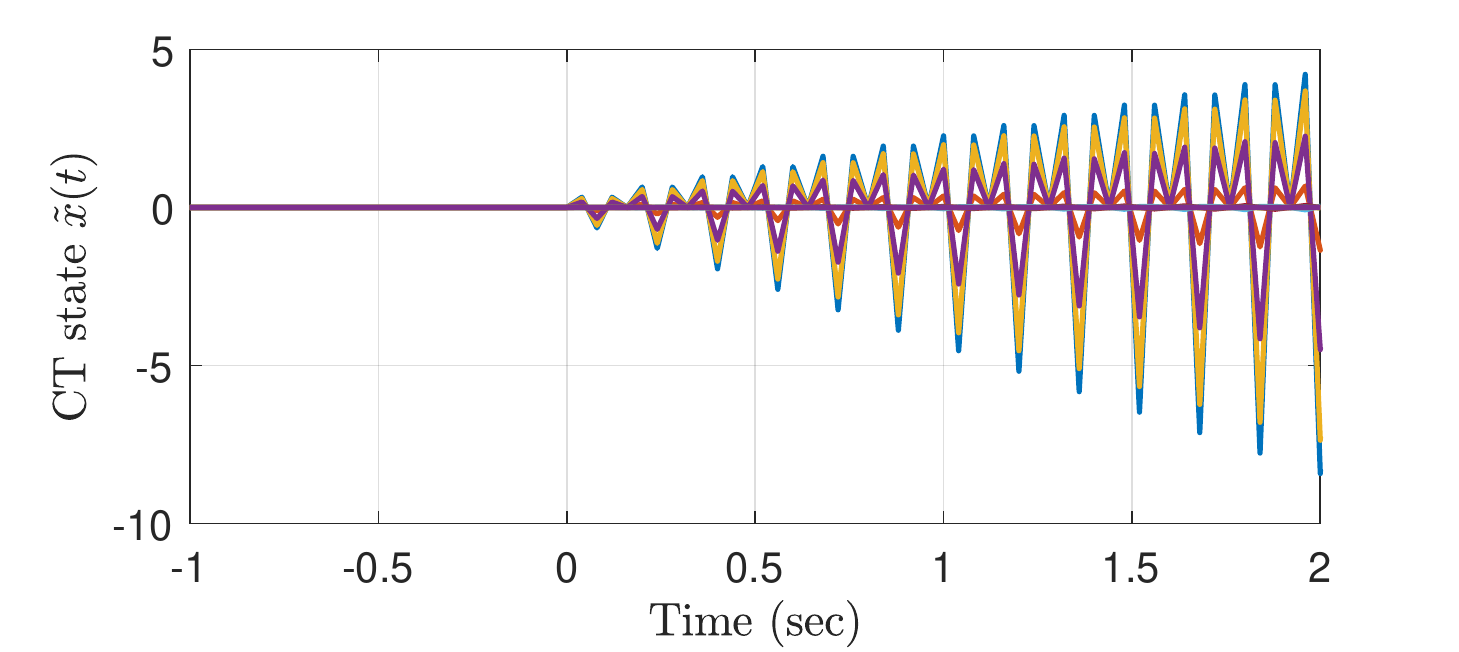}
	\end{center}
	\caption{Continuous-time error state $\tilde{x}(t)$ of the X-38 model}\label{fig: state}
\end{figure}

\begin{figure}[t!]
	\begin{center}
		\subfigure[Continuous-time output $y_{\mathsf{o}}(t)$ under no attack]{\includegraphics[width=0.5\textwidth]{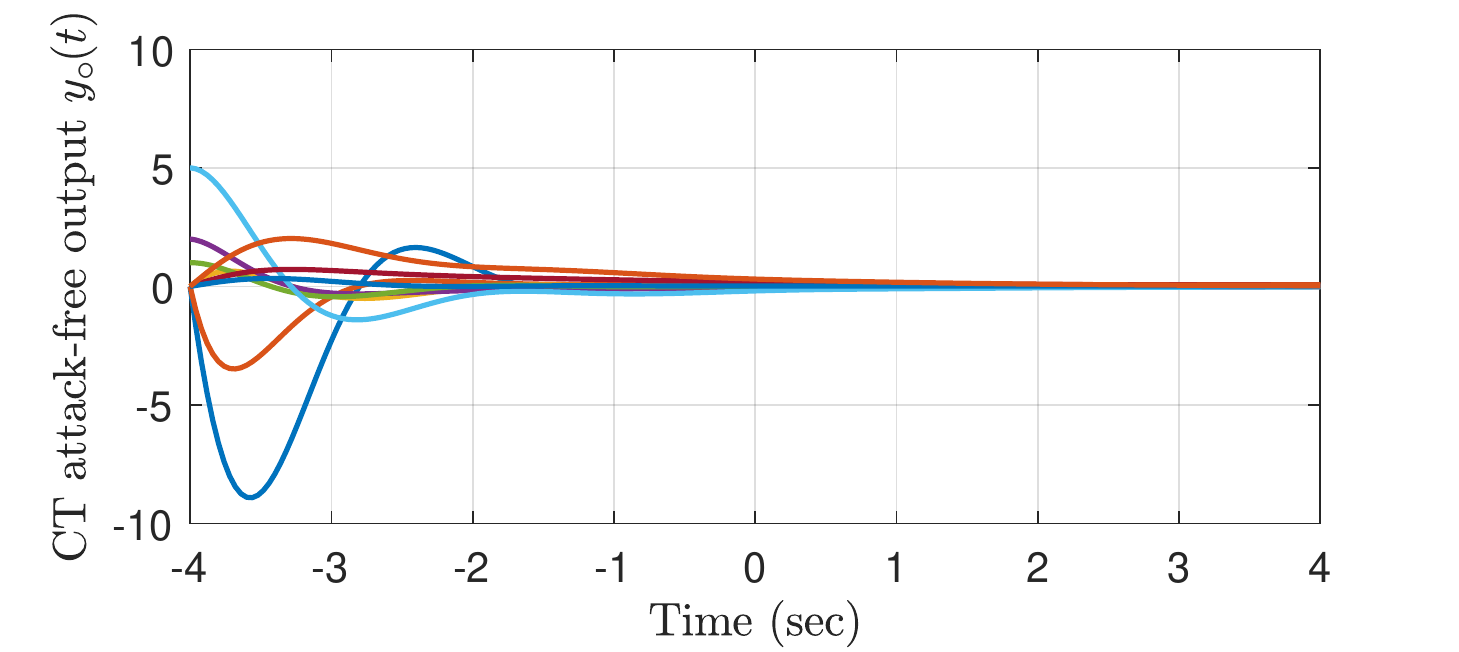}}
		\subfigure[Continuous-time output $y(t)$ under the proposed attack]{\includegraphics[width=0.5\textwidth]{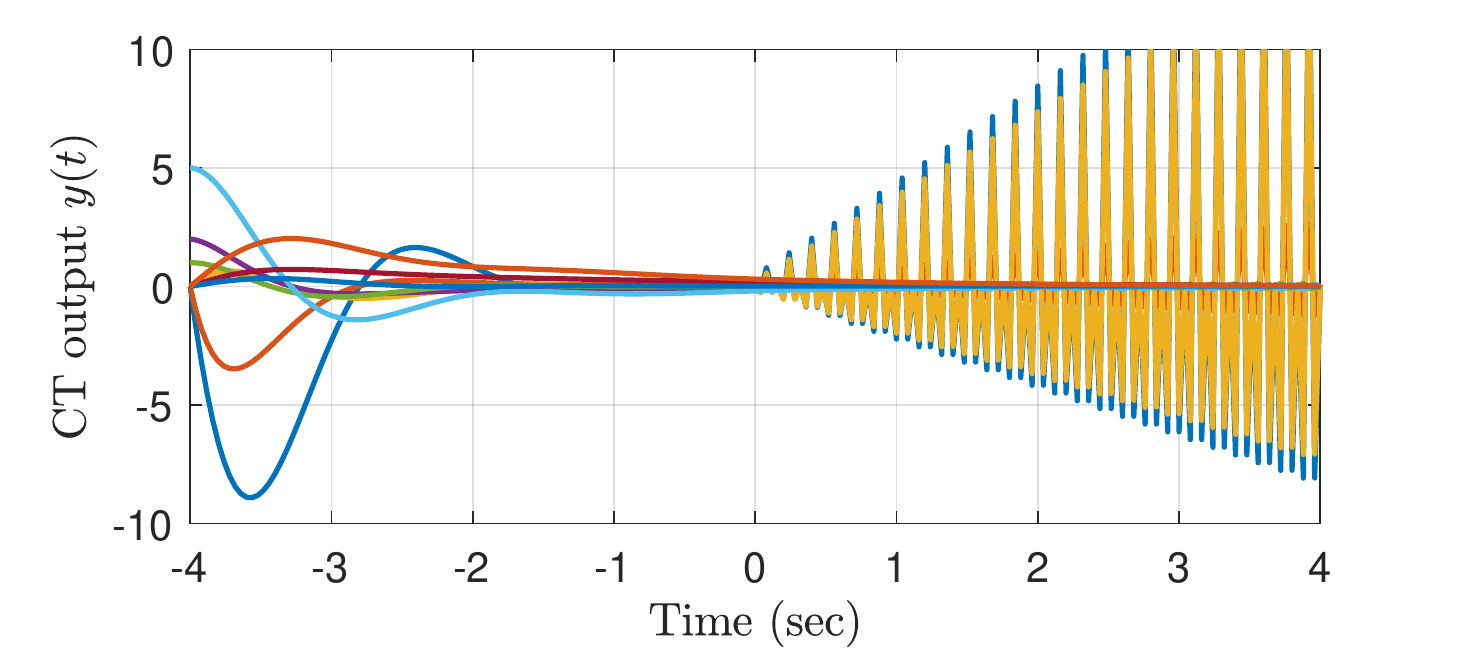}}
		\subfigure[Discrete-time output $y(jT_s)$ under the proposed attack]{\includegraphics[width=0.5\textwidth]{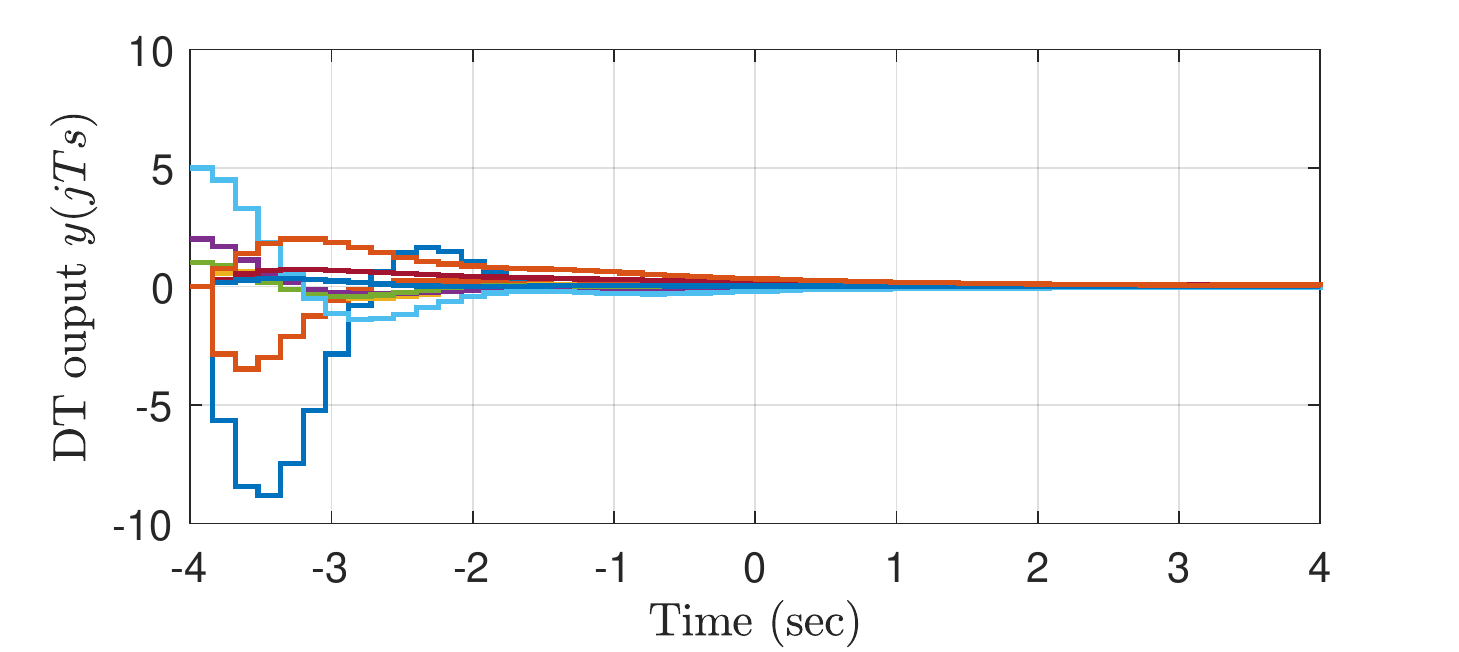}}
	\end{center}
	\caption{Continuous- and discrete-time outputs with and without attack}\label{fig: output}
\end{figure}

To see the effect of the attack, the attack sequence is injected into the input channel of the plant at $t=0~{\rm sec}$.
Fig.~\ref{fig: attack} shows the injected attack and Fig.~\ref{fig: state} illustrates the state error $\tilde x(t)$.
In spite of the disruptive property seen in Fig.~\ref{fig: state}, the measured output at sampling times look normal (Fig.~\ref{fig: output}.(c)).
In fact, the continuous-time output $y(t)$ is not calm as seen in Fig.~\ref{fig: output}.(b) while the attack-free continuout-time output $y_{\mathsf{o}}(t)$ is also depicted for comparison in Fig.~\ref{fig: output}.(a).

\subsection{Numerical Example: $R=0.4$ with $\delta = 0.75$}\label{sec:N<1example}

In this subsection, we show that the proposed attack is effective under Assumption \ref{asm: input redundancy}, even if the sampling period of the sensor is shorter than that of the actuator. 
Let us consider the case where $T_a=1~{\rm sec}$ and $T_s=0.4~{\rm sec}$, so that $R = 0.4/1 = 2/5 = \beta/\alpha$ (i.e., there are $5$ sensings and $2$ actuations for each cluster). 
Moreover, let us assume an offset with $\delta = 0.3/0.4 = 0.75$ (i.e., the sensor starts $0.3~{\rm sec}$ later than the actuator). 
The considered plant is described by a minimal realization of  
$$G(s)=\begin{bmatrix}
\dfrac{1}{s+1}& \dfrac{2}{(s+2)(s+3)} & \dfrac{4}{(s+4)(s+5)}
\end{bmatrix}.$$
From the minimal realization $A \in \R^{5 \times 5}$, $B \in \R^{5 \times 3}$, and $C \in \R^{1 \times 5}$, one can verify Assumption \ref{asm: input redundancy} as follows:
\begin{itemize}
	\item The plant has $3$ inputs, $1$ output, and $R=0.4$. Hence, $q=1 < Rp=1.2$, and so, the condition (a) holds (i.e., $\mathcal{C}\Pi \in \R^{5 \times 6}$ so that $\ker \mathcal{C}\Pi \neq \{0\}$).
	\item The matrix $B_\ddd$ has full column rank, and there exists a non-zero vector $z$ such that $\mathcal{C}\Pi z = 0$ where the first 3 components are a non-zero vector in $\R^3$. Then, by the proof of Proposition \ref{prop:b}, $i^*=1$. Therefore, the condition (b) holds with $t_k^* = i^*/\beta = 1/2$.
	\item The matrix $\mathcal{C}\Pi$ has full row rank so that ${\rm im}~\mathcal{C}\Pi=\mathbb{R}^5$ and the condition (c) holds.
\end{itemize}

An attack sequence $\bar a[i]$ is constructed as proposed with $H_k = 10k$.
In particular, we have chosen $\eta_{\langle k \rangle} = \eta =  {\rm col}(-0.188, -0.163, 0.746, 0.138, -0.467, 0.379)  \in \ker\mathcal{C}\Pi$ for all $k \ge 1$, which satisfies $\Phi_k^* \eta \not = 0$.
The quantities $\kappa_k$ and $\zeta_{\langle k \rangle}$ are selected appropriately by the attack generation procedure in Section \ref{sec: main result}.

\begin{figure}[t!]
	\begin{center}
		\includegraphics[width=0.5\textwidth]{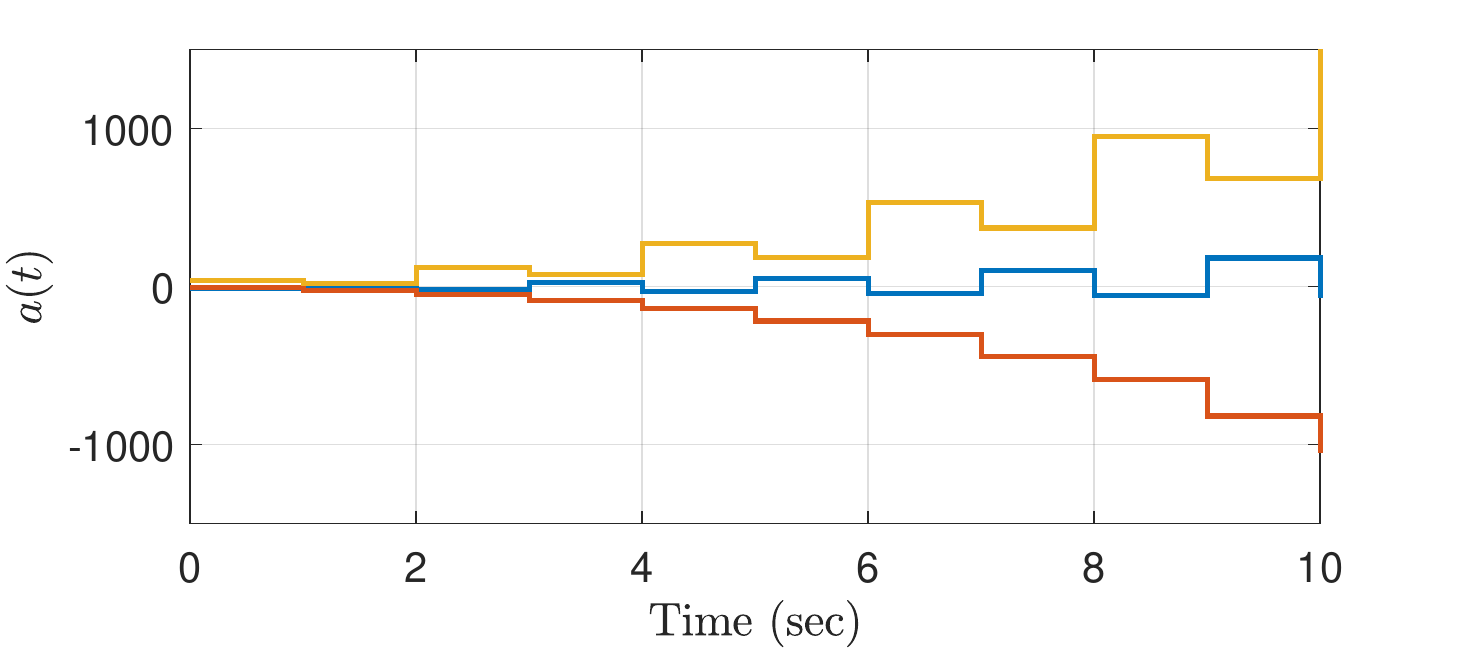}
	\end{center}
	\caption{Attack sequence $ a (t)$}\label{fig:attack_N<1}
\end{figure}

\begin{figure}[t!]
	\begin{center}
		\includegraphics[width=0.5\textwidth]{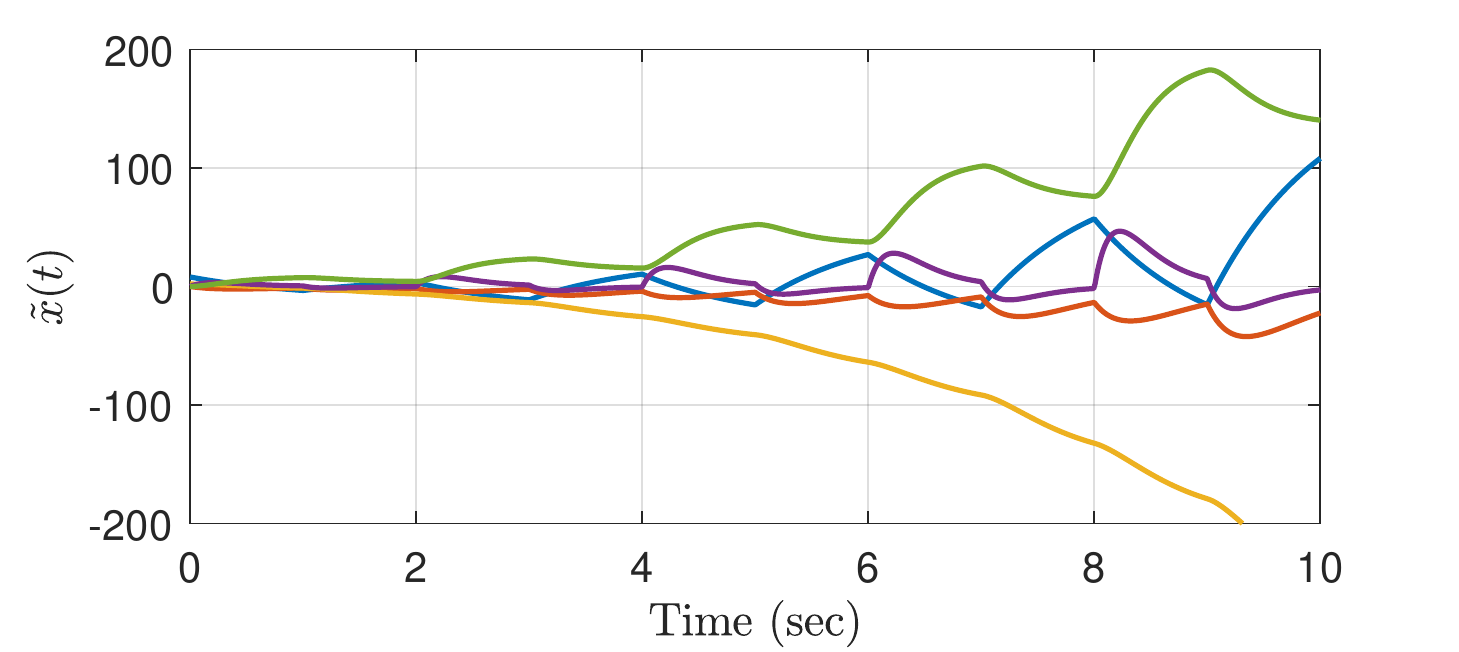}
	\end{center}
	\caption{Behavior $\tilde x(t)$ of error dynamics}\label{fig:tilde_x_N<1}
\end{figure}

\begin{figure}[t!]
	\begin{center}
		\includegraphics[width=0.5\textwidth]{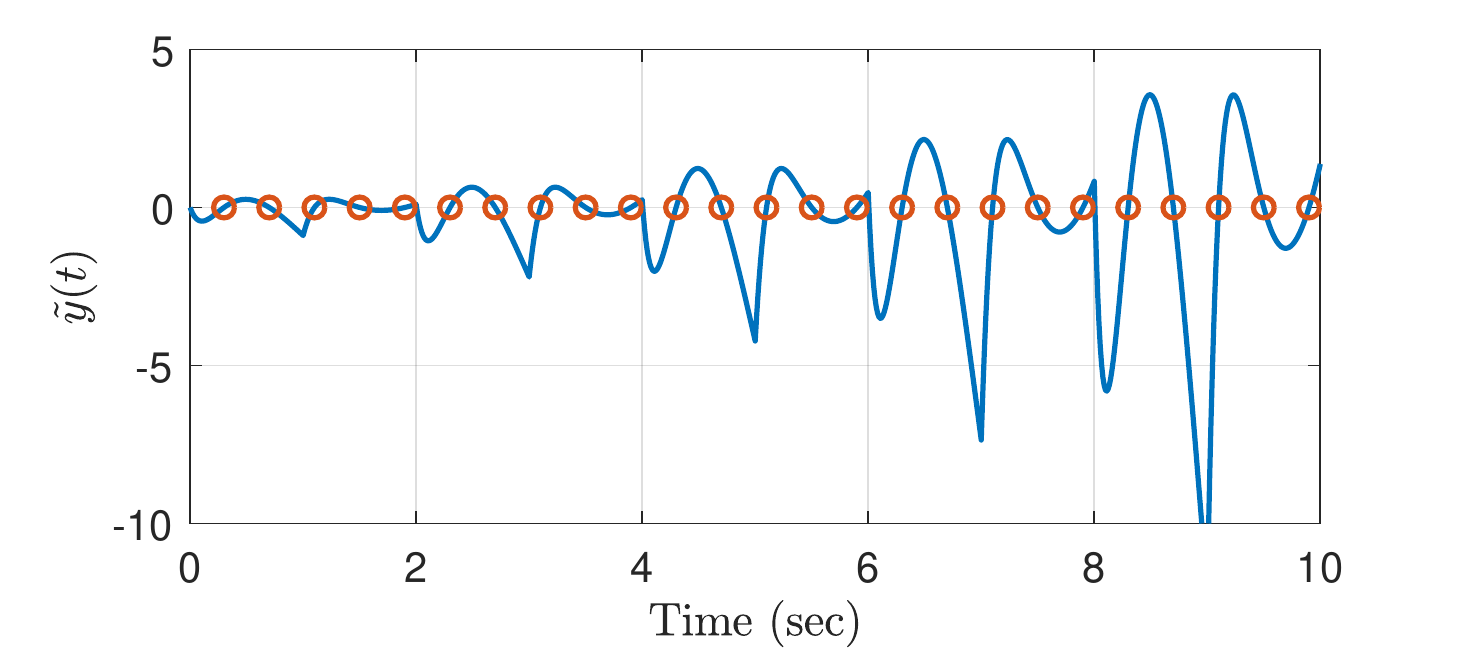}
	\end{center}
	\caption{Output $\tilde y(t)$ (blue solid line) and its sampled measurements $\tilde y(j_\delta T_s)$ with offset $\delta = 0.75$ (red circle)}\label{fig:tilde_y_N<1}
\end{figure}

The simulation results illustrate the constructed attack signal in Fig.~\ref{fig:attack_N<1}, the behavior of $\tilde x(t)$ in Fig.~\ref{fig:tilde_x_N<1}, and the output signal $\tilde y(t)$ in Fig.~\ref{fig:tilde_y_N<1}, respectively.
It is seen that, even if the error variable $\tilde x(t)$ and the continuous-time output $\tilde y(t)$ diverge, the output measurements (represented as red circles in Fig.~\ref{fig:tilde_y_N<1}) remain zero, so that both stealthiness and disruptive property are achieved.

\begin{figure}[t!]
	\begin{center}
		\includegraphics[width=0.5\textwidth]{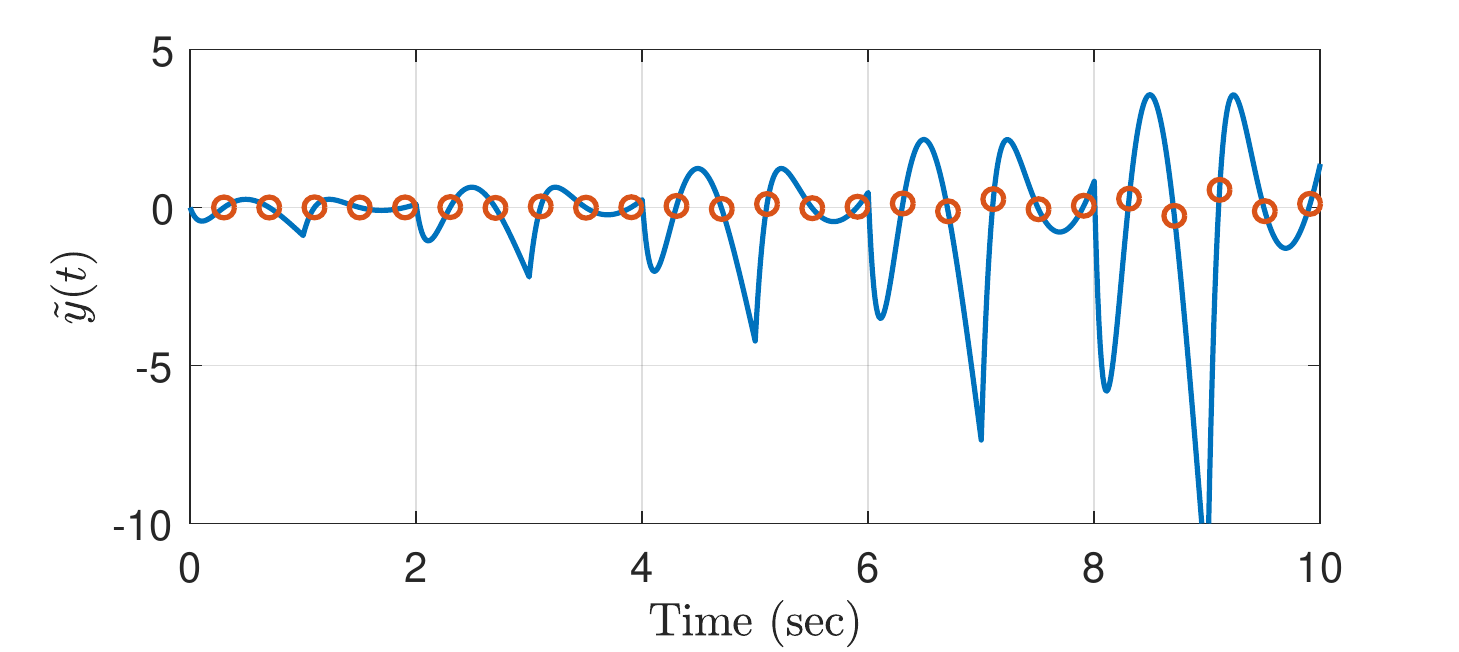}
	\end{center}
	\caption{Output $\tilde y(t)$ and its sampled measurements when $R=0.4004$, $T_s = 0.4004$ and $T_a = 1$ while the attack is designed assuming $R=0.4$, $T_s=0.4$, and $T_a=1$.}\label{fig:realR}
\end{figure}

Out of curiosity, we have simulated the case where the actual $R$ is $0.4004$ but is estimated as $R=0.4$, so that the attack signal is designed based on $R=0.4$.
As seen in Fig.~\ref{fig:realR}, the measured output does not remain zero forever, but if the estimate is sufficiently close to the true value, it is expected that the detection of the attack is delayed until a fatal damage is incurred in the plant.

Finally, in order to detect such an attack, one may deploy a mechanism of intermittent output sampling in addition to periodic sampling. Clearly, Fig. \ref{fig:tilde_y_N<1} and \ref{fig:realR} show that additional output sample will yield non-zero values that would call attention of the operator.
	
\section{Concluding Remark and Future Works} \label{sec: conclusion}

It has been recently studied that the interconnection between continuous- and discrete-time components may make some CPS more vulnerable to cyber-physical attacks (for instance, the zero-dynamics attack targeting the sampling zeros \cite{MNP15,Kim17}). 
We have clarified in this paper that another type of zero-stealthy attack is also possible, if there exists enough input redundancy for the system in the multi-rate or multi-input sense. 
By taking a closer look at the state trajectory in both continuous- and discrete-time domains, we showed that how the additional input resources and full system knowledge enable the adversary to compromise the inter-sample behavior of the sampled-data system, while being perfectly undetected at each sampling time. 

Future works include consideration of input saturation and  investigation of the case when $R$, the ratio of $T_s$ and $T_a$, is real number.
By analyzing the proposed construction of attack signal, we expect to figure out quantitative relationship between $H_k$ and the saturation level.
When the ratio between sampling period and actuation period is a real number, by approximating it as a rational number sufficiently closely, we expect to delay the detection of attack as much as we want, under strictly positive error threshold of anomaly detector.

Finally, it is also necessary to develop a method to detect the proposed attack.
At this moment, we just think that intermittent random sampling of the output in addition to periodic sampling, removing unnecessary input channels, or concealing the system knowledge may be helpful.

   \begin{IEEEbiographynophoto}{Jihan Kim}
   	received his B.S. degree in the School of Electronic Engineering from Sogang University in 2014.
   	Since 2014, he has been working toward his Ph.D. degree at Seoul National University.
   	His research interests include security of cyber-physical systems, and sampled-data system.
   \end{IEEEbiographynophoto}

   \begin{IEEEbiographynophoto}{Gyunghoon Park}
   	received his B.S. degree in the School of Electrical and Computer Engineering from Sungkyunkwan University in 2011, and M.S. degree from the School of Electrical Engineering and Computer Science, Seoul National
   	University in 2013, respectively. Since 2013, he has been working toward his Ph.D. degree at Seoul National University.
   	His research interests include theory and application of disturbance observer, security of cyber-physical systems, and sampled-data system.
   \end{IEEEbiographynophoto}
      
   \begin{IEEEbiographynophoto}{Hyungbo Shim}
   	received the B.S., M.S., and Ph.D. degrees from Seoul National University, Korea, and held the post-doc position at University of California, Santa Barbara till 2001. 
   	He joined Hanyang University, Seoul, Korea, in 2002. 
   	Since 2003, he has been with Seoul National University, Korea. 
   	He served as associate editor for Automatica, IEEE Trans. on Automatic Control, Int. Journal of Robust and Nonlinear Control, and European Journal of Control, and as editor for Int. Journal of Control, Automation, and Systems. 
   	He was the Program Chair of ICCAS 2014 and Vice-program Chair of IFAC World Congress 2008. 
   	His research interest includes stability analysis of nonlinear systems, observer design, disturbance observer technique, secure control systems, and synchronization.
   \end{IEEEbiographynophoto}
   
   \begin{IEEEbiographynophoto}{Yongsoon Eun}
   	(M'03) received the B.A. degree in
   	mathematics, and the B.S. and M.S.E. degrees in
   	control and instrumentation engineering from Seoul
   	National University, Seoul, Korea, in 1992, 1994,
   	and 1997, respectively, and the Ph.D. degree in electrical
   	engineering and computer science from the
   	University of Michigan, Ann Arbor, MI, USA, in
   	2003.
   	From 2003 to 2012, he was a Research Scientist
   	with the Xerox Innovation Group, Webster, NY, USA,
   	where he worked on a number of subsystem technologies
   	in the xerographic marking process and image registration method in inkjet marking technology. 
   	Since 2012, he is an Associate Professor with the Department of Information and Communication Engineering, Daegu Gyeongbuk Institute of Science and Technology (DGIST), Daegu, Korea. 
   	His research interests include control systems with nonlinear sensors and actuators, networked control systems, cyber-physical systems, and resilient control systems.   
   \end{IEEEbiographynophoto}

\end{document}